**Astro2020 Science White Paper**

# Characterizing the Distribution of Parameters of Planets Found by Radial Velocity is Essential for Understanding Planet Formation and Evolution

**Thematic Areas:** 1. ☒ Planetary Systems   2. ☒ Star and Planet Formation
3. ☒Stars and Stellar Evolution

**Principal Author:**
Name:          Stuart F. Taylor
Institution:   Participation Worldscope
Email:         astrostuart@gmail.com
Phone:         +852 9847-0748

**Abstract**:
Features in the distribution of exoplanet parameters by period demonstrate that the distribution of planet parameters is rich with information that can provide essential guidance to understanding planet histories. Structure has been found in the counts of planet-star ``objects'' by period, and within these structures, there are different correlations of eccentricity with planet number, stellar metallicity, planet count density per log period, stellar multiplicity, and planet mass. These appear to change with each other, and with stellar mass, but there are too few planets to easily and reliably study these important relationships. These relationships are the bulk observables against which the theory of planet formation and evolution must be tested. The opportunity to determine the nature of the relationships of the exoplanet parameters on each other demonstrate the value of finding more planets across period ranging from days to thousands of days and beyond. We recommend support for continuing to find more planets, even giant planets, with periods up to periods of a few thousand days. We also recommend support for the study of the distribution of the many exoplanet parameters.

[2019 August note added to arxiv version:
Publication citation and updated reference appended to end.]

## Introduction and motivation

The best indicators of planet formation and evolution must include the distribution of exoplanets parameters by period. Measurables such as the density of planet counts and the relationships of eccentricity with other parameters can show the results of how planets form and evolve.

Studying the distribution of parameters requires finding and measuring the parameters of more planets. We advocate the value of continued work using radial velocity (RV) observations to search for planets in the period range of hundreds to over one thousand days in order to better characterize the ``main pileup'' of planets found in this range. We present some of the features in the main pileup of planets found by RV, but we use how we found unexpected features in this region to advocate the study of the distribution of exoplanets in general.

While we show an unexpected level of features in the current data set, we show that the current data set is too small to adequately evaluate the dependence of these features on other important parameters.

## An emerging picture requires more data

Patterns providing what could be emerging evidence for the picture of giant planets forming in a narrow ring in systems with stars of metallicities below solar, but for this ring becoming split into two planet forming regions for planets of stars of solar and higher metallicities.

## Features in the main pileup.

To understand the history of how planets systems evolve it is essential to study the main pileup of planets. It has become even more important to understand the distribution of the parameters of planets because of some significant features discovered in the distribution of the main populations of planet in the highest populations region of planets, for which we here define the region of interest (ROI) of the "main pileup" of planets to go from periods of 100 to 5000 days. The finding that the counts of planets increases in the several hundred day period range has been reported by (U07; W09; HP12; BN13).

The period range from 100 to 5000 d brackets this pileup, and contains 313 of the 434 objects found by RV before 2016 that we study. These features show the importance of continuing long-term RV observations to increase the numbers of planetary objects found in this region, given how the current number has been enough to identify that these features exist, but it is too small to reliably study the dependence of these features on the set of several parameters of the planet-orbit-star system that we call an ``object.'' It is worrisome that fewer long period planets have been reported in the last few years. We advocate that the study of the full largest pileup of planets, and that RV surveys continue observations aimed at characterizing the planet populations at periods continuing up to the 5000 days of the current data set.

**Attributes of Features Needing Better Statistics**

Presented here are correlations of eccentricity with five parameters among the 188 sets of parameter data, or "objects", associated with those planets hosted by sunlike stars ("SL", where stellar log g > 4 and 4500 < Teff < 6500), from among the 434 planets found by the radial velocity (RV) method with periods from 100 to 5000d before 2016. The mean eccentricity of the 34 planets of stars with stellar companions, or SLBS objects, is 0.438 which is higher than the mean eccentricity of 0.288 for the 154 SLSS objects. Though this likely does represent the influence of binary stars, there are other potentially confounding correlations that makes it important to compare various subpopulations, but it we degrade our statistics to divide the 34 SLBS objects further. It is particularly difficult to separately study the 34 SLBS objects by metallicity when there are 27 rSLBS but only 7 pSLBS objects.

The 154 SLSS objects can be divided into data-set objects representing 113 and 41 planets of metal-rich and poor stars, or rSLSS and pSLSS objects, respectively. T19 shows these 113 are enough to establish the presence of the two peaks and gap among the log period distribution of rSLSS objects, Fig. 1. The white paper T18 presented the features of this peak-gap-peak feature of the rSLSS population. [Update 2019 August: This peak-gap-peak features are presented in detail in **T19**.]

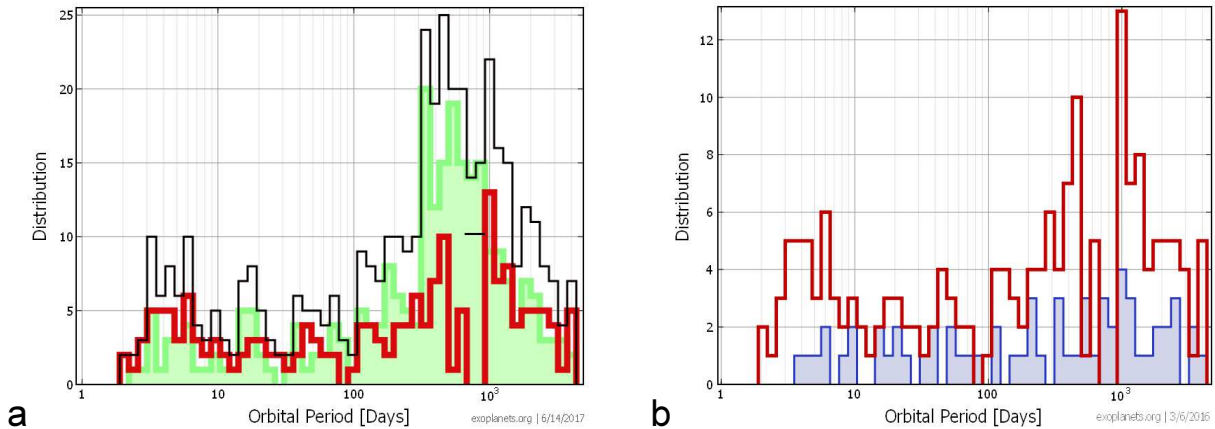

Fig. 1. 1a (Left): A histogram of all objects (black) shows the main pileup peaking at periods of between 300 and less than 1000 days but have a notch giving a hint of bimodal constituent. Also shown are the rSLSS objects (red) which have the bimodal peak, and the sum of all others (green), which have a single peak. The width of the deep gap (with zero objects) in log period space is shown as a line. 1b (Right): The pSLSS (blue) are shown with the rSLSS objects (red).

The presence of these features raises questions of how do these features depend on other parameters, such as the number of planets in the system, or the mass of the star. To consider whether these features exist in both single-planet and multiplanet systems, among the 113 objects we find 63 single planet objects and 50 multiplanet objects. A comparison of histograms of the two gaps and peaks with only single and multiple planet systems with the distribution for all 113 object in Fig. 1 make it appear fairly certain that the peak-gap-peak pattern is a feature of planets in single planet systems, but for multiplanet systems it is more difficult to use just 50 objects to say whether both of the two peaks exist, and among these the presence of the gap is less certain. While the multiplanet selection contains a spike of objects at the onset of the long period pileup of perhaps 10 objects that leads us to suggest that there is a similar though perhaps smaller peak at 923 day periods, there is only a low significance rise in the numbers of objects at the short period peak. While there is also an absence of objects in the region of the deep part of the gap, the mean density of objects is too low to be sure that we do not see the gap in among the 50 objects in the multiplanet selection of rSLSS objects. The only way to answer whether planets in multiple planet systems display these features is to find more such planets.

The low number of objects also makes it difficult to consider whether these features might depend on stellar mass. When the 113 rSLSS objects are separated by whether the stellar mass is less or more than that of the sun, there are 40 low and 73 high stellar-mass objects. Though both of the peaks and the gaps appear in both selections, there are tantalizing possible differences in the widths and the relative heights of the peaks. It appears that in the higher stellar mass selection that both peaks may be wider, i.e. the short period peak may start at even shorter periods and the longer period peak may extend to longer periods, but without better statistics these possible features remain uncertain versus either random events. It will take finding more longer period objects of low stellar mass to know if the more narrow width of the longer period peak is simply due to it being more difficult to detect radial velocities with longer periods exhibited by planet-hosting lower mass stars.

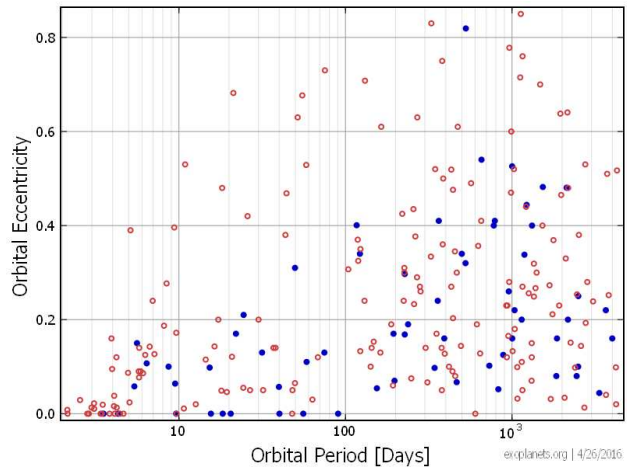

Fig. 2. The eccentricity as a function of period for sunlike stars, found by radial velocity, with filled blue circles for objects more iron-poor than the sun, and open red circles for objects more iron-rich. The mean eccentricity tends higher near where the planet count density is highest.

**Eccentricity correlations with five parameters as a function of period:**

The 2018 white paper T18 shows that eccentricity likely has a correlation (or similar relationship) with five other parameters among SL objects, relationships that typically change with period: number of planets in system (L&T 2015), eccentricity (T12, DM13, T13), whether the star is binary (T13), planet mass (A13, W09, U07). Current work is showing that eccentricity appears correlated with the aggregate density of planets per log period (T19) when separately considering the 113 rSLSS and 41 pSLSS objects.

The eccentricity by period, separated by metallicity, is shown in Fig. 2. The relations of eccentricity with metallicity by period may be a manifestation of a correlation of eccentricity with the aggregate counts per log period density of planets, which changes with metallicity. This shown as the double-peak feature for the 113 rSLSS objects in Fig. 1, but the 41 pSLSS objects have a single pileup, Fig. 1b.

While these have been enough to see likely correlations, the 113 rSLSS and 41 pSLSS objects are too small numbers of objects to study details of these correlations. Indeed, the major features are not easily proven.

### Shorter period (less than ~100 d) eccentricity-metallicity correlation

Eccentricity is correlated with metallicity in shorter periods that described above (DM13, T12b, T13b). Several short period features make the short period region worthy of a separate report. We note the white paper T13c advocating finding more giant planets to obtain data on the location of the shortest period falloff with stellar age, to give data on the strength of tidal dissipation in the star (1/Q). Shortest period giant planets should show period decreases that depend on a combination of the strength of stellar tidal migration and the rate of inward planet migration. When evaluating the strength of stellar tidal migration, it is essential to consider how a very small rate of planet migration inwards into the star can change the results dramatically (T12a, T12b, T13a). Being able to observe a period decrease in the shortest period planets would provide a separate measurement of stellar tidal dissipation that would remove this degeneracy.

### Objective: Finding what might cause the gaps and peaks.

For these patterns to show up from so many different systems shows that planet formation and evolution must occur much more uniformity than previously assumed. While it is not surprising that there is an increase in counts corresponding to an important condensation temperature, the presence of such a deep gap at periods shortward of the main peak, with another shorter period pileup, represents a new major unexpected feature of the distribution of a major fraction of exoplanetary systems.

*Recent Status of RV searches:* There has been a slowing down of RV-based discoveries finding lower mass planets, with the number of sunlike objects with periods from 100 to 5000 d only increasing by seven going from the datasets of before 2016 to before 2019. This prevents exploring how these features depend on planet mass. An important question is whether the gap in log period of ``rSLSS'' planets continues down to lower masses than the smallest masses available in the 2016 dataset.

*Recommendation: Support finding more planets seeking lower masses by continuing searches using RV. This is needed to overcome the following challenges of limited data:* There are only 434 objects with periods from 100 d to 5000d found by RV before 2016, with only a handful more having been found since then. In fact, the pace of discovering more objects has begun declining as more resources to perform RV measurements is diverted to measuring planet candidates found by other methods.

## References

[T19, updated reference that includes paper accepted by Astronomiche Nachrichten at end]

**Linked Reference Material:**

T19 refers to the following ResearchGate.net project that includes the draft paper describing the double-peak gap features and an explanation of how to plot it:
https://www.researchgate.net/project/Rings-and-Gaps-Everywhere-Features-in-the-distribution-of-planets

This project contains a draft of a paper in preparation. The link includes the following:

An explanation showing how the double-peak gap features are extremely difficult (on the order of many $10^4$) to result from random distributions of observations, and would be very difficult to result from observational effects.

A "How-To'' guide to quickly plot the double-peak-gap feature on exoplanets.org.

**Acknowledgements:** Data for this work taken from the Exoplanet Data Explorer at exoplanets.org.

**Appended 2019 August 7:**

Published as

Taylor, Stuart F., Astro2020: Decadal Survey on Astronomy and Astrophysics, science white papers, no. 179; Bulletin of the American Astronomical Society, Vol. 51, Issue 3, id. 179 (2019); **2019 BAAS 51(3) 179**

[Update since 2019 March submission to BAAS, 2019 August 7: Paper on double-peak with gap pattern in counts accepted by AN; reference appended to end.]

**Updated reference for T19:**

T19: Taylor, Stuart F., 2019, "Rings and Gaps Everywhere?: Features in the distribution of planets", Accepted for publication in Astronomiche Nachrichten.

As of 2019 August, this is available in the Researchgate.net project listed above at the top of the references.

The direct link to this reference is at:

https://www.researchgate.net/publication/335016168_Unexpected_gap_creating_two_p eaks_in_the_periods_of_planets_of_metal-rich_sunlike_single_stars

T19 has also being listed on the arxiv preprint server that has been listed as 1908.01679, though the identifier may be changed.